\documentclass[useAMS,usenatbib]{mnras}
\usepackage{graphicx,natbib,amssymb,amsmath,stmaryrd,makecell}
\usepackage{txfonts}
\usepackage{xcolor}
\usepackage{hyperref}
\usepackage{xspace}
\usepackage{caption}
\usepackage{lipsum}
\usepackage{soul}
\usepackage{bm}

\newcommand{\GHz}{{\rm GHz}}
\newcommand{\MHz}{{\rm MHz}}

\newcommand{\expf}[1]{{{\rm e}^{#1}}}
\newcommand{\Jbb}{\mathcal{J}_{\rm bb}}
\newcommand{\Jdc}{\mathcal{J}_{\rm DC}}

\newcommand{\TCMB}{T_{\rm CMB}}

\newcommand{\zh}{{z_{\rm h}}}

\newcommand{\Planck}{{\it Planck}\xspace}

\newcommand{\xe}{x_{\rm e}}

\newcommand{\xc}{x_{\rm c}}
\newcommand{\Xe}{X_{\rm e}}

\newcommand{\id}{{\,\rm d}}

\newcommand{\beq}{\begin{equation}}   %

\newcommand{\eeq}{\end{equation}}   %

\newcommand{\beqa}{\begin{eqnarray}}   %

\newcommand{\eeqa}{\end{eqnarray}}   %

\newcommand{\beal}{\begin{align}}

\newcommand{\enal}{\end{align}}

\newcommand{\bspl}{\begin{split}}

\newcommand{\espl}{\end{split}}

\newcommand{\bsub}{\begin{subequations}}

\newcommand{\esub}{\end{subequations}}

\newcommand{\bmulti}{\begin{multline}}   %

\newcommand{\beqm}{\begin{mathletters}}   %

\newcommand{\eeqm}{\end{mathletters}}   %

\newcommand{\kB}{k_{\rm B}}

\newcommand{\me}{m_{\rm e}}

\newcommand{\Ne}{N_{\rm e}}

\newcommand{\Te}{T_{\rm e}}

\newcommand{\Tg}{T_{\gamma}}

\newcommand{\Thg}{\theta_{\gamma}}

\newcommand{\sigT}{\sigma_{\rm T}}

\newcommand{\pot}[2]{#1 \times 10^{#2}}

\newcommand{\Imu}{\mathcal{I}_{\hat{\mu}}}
\newcommand{\Drr}{\frac{\Delta \rho_{\gamma}}{\rho_{\gamma}}}

\newcommand{\COBEF}{{\it COBE/FIRAS}\xspace}

\newcommand{\PIXIE}{{\it PIXIE}\xspace}
\newcommand{\FOSSIL}{{\it FOSSIL}\xspace}
\newcommand{\WMAP}{{\it WMAP}\xspace}

\newcommand{\Jdiss}{\mathcal{J}_{\rm d}}

\author[Evangelista, Chluba \& Pace]{ 
Sara Evangelista$^{1,2}$\thanks{E-mail:sara.evangelista887@edu.unito.it}, Jens Chluba$^2$ and Francesco Pace$^{1,3,4}$
\\
$^1$Dipartimento di Fisica, Universit\`a degli Studi di Torino, Via P. Giuria 1, I-10125, Torino, Italy
\\
$^2$Jodrell Bank Centre for Astrophysics, School of Physics and Astronomy, The University of Manchester, Manchester M13 9PL, U.K.\\
$^3$ INFN-Sezione di Torino, Via P. Giuria 1, I-10125, Torino, Italy\\
$^4$ INAF-Istituto Nazionale di Astrofisica, Osservatorio Astrofisico di Torino, strada Osservatorio 20, 10025, Pino Torinese, Italy
\vspace{-2mm}
}

\date{Accepted XXX. Received YYY; in original form ZZZ}

\title[Late-time heating Green's]
{The late-time heating Green's function and improvements to distortion frequency hierarchy treatment}

\begin{document}

\maketitle

\begin{abstract}
Early energy injection leaves an imprint on the observed blackbody spectrum of the CMB, allowing us to study the thermal history of the Universe. For small energy release, the distortion can be efficiently computed using the quasi-exact Green's function method. For pre-recombination injections, the Green's function has already been studied previously. Here we reconsider the pre- and post-recombination periods, showcasing both the spectral distortion intensity and the relative temperature difference, which encrypt precious information about physical processes such as free-free interactions and thermal decoupling. We present the associated distortion visibility function, investigating the impact of various physical effects. 
We then study improvements to the so-called frequency hierarchy (FH) treatment, a method that was developed for the modelling of anisotropic distortions, which like the average distortion signals encode valuable cosmological information. Specifically, the FH treatment has shortcomings even in the $\mu$ era, that in principle should be easy to overcome. In this paper, we introduce a new approach to reduce the mismatch, concluding with a redefinition of the $\mu$ spectral shape using {\tt CosmoTherm}. This solution takes into account double Compton and Bremsstrahlung effects in the low tail, which can be included in the FH. This opens the path towards a refined modeling of spectral distortion anisotropies.
\end{abstract}

\section{Introduction}
\label{sec:Intro}
CMB spectral distortions have been recently recognized as a fundamental probe to constrain early Universe physics, complementing the studies on temperature and polarization anisotropies. Measurements of temperature anisotropies performed by \WMAP \citep{Komatsu2010} and \Planck \citep{Planck2018params}, were able to precisely determine key cosmological parameters, such as the curvature of the Universe and the dark matter density, as well as provide firm evidence for an inflationary Big Bang model. On the other hand, future measurements of polarization $B$ modes, related to primordial gravitational waves, promise to put even tighter constraints on inflation \citep{Kamionkowski1997, Seljak1997}.

However, if we are interested in processes that occurred before recombination ($z\simeq 1000$), one can also look at the distortions of the CMB spectrum \citep{Sunyaev1970mu, Burigana1991, Hu1993}. More than 30 years ago, the \COBEF space-based experiment \citep{Fixsen1996, Fixsen2003} measured the CMB blackbody at a temperature of $T_{\rm CMB,0} = 2.725 \pm 0.002 {\rm K}$ with unprecedented precision, providing an upper limit on departures of $\Delta I/I \lesssim 10^{-5}$. The limits of \COBEF still stand, but the great potential of CMB spectral distortion science has led to a renewed interest in the field, opening the door to theoretical work as well as to new experiments \citep{Chluba2021Voyage}. In particular, the balloon-borne spectrometer BISOU \citep{BISOU} has recently entered Phase A, preparing to measure $y$-distortions with a sensitivity $\simeq 20-30$ higher than \COBEF, giving a taste of what could be achieved with future space missions like \FOSSIL or \PIXIE\citep{Kogut2011PIXIE, Kogut2016SPIE, Kogut2024}. Meanwhile, ground-based experiments such as TMS \citep{Jose2020TMS}, ASPERa \citep{Mayuri2015} and COSMO \citep{Masi2021} are making progress, with first measurements of TMS at $10-20\,\GHz$ expected soon. 

Spectral distortions (SDs) are formed by out-of-equilibrium processes, which modify the blackbody of the CMB, creating a spectral distortion. Their shape depends on the redshift of injection, enabling us to constrain primordial processes deep into the pre-recombination Universe. For heating redshifts $5 \times 10^4 \lesssim \zh \lesssim 2 \times 10^6$, a $\mu$ distortion is produced, which gradually morphs toward a $y$ distortion at $\zh\lesssim 5 \times 10^4$. SDs thermalise due to Compton scattering and photon emission/absorption processes (double Compton scattering and Bremsstrahlung), reducing significantly their amplitude in agreement with the observational limits reported above. 

If the energy release rate is known, the resulting distortion can be numerically computed by solving the full Boltzmann equation. Although this can be computationally challenging, the problem can be further simplified by implementing the Green's method in {\tt CosmoTherm} \citep{Chluba2013Green, Chluba2015GreensII}. The definition of the Green's function is in turn related to the distortion visibility function, which describes how much of the injected energy has thermalized. Both the SD and the distortion visibility function can be affected by various physical processes at late times, such as the Hubble cooling effect after thermal decoupling of electrons and photons and the fraction of heating which goes into the CMB, to name a few that will be investigated here.

So far, the focus has been primarily on average distortions science, but spatial information can be equally helpful for studying new physics posing novel challenges to the field. Nevertheless, SD anisotropies are smaller and therefore more challenging to observe and, at the same time, their computation is extremely difficult for existing Boltzmann codes. Due to the number of $k$ modes that have to be considered, the number of equations that have to be solved rises to approximately $ 10^3 - 10^4$. The Frequency Hierarchy (FH) treatment \citep{chluba_spectro-spatial_2023-I, chluba_spectro-spatial_2023-II} is a recently developed spectral discretisation that reduces the complexity of the system by two orders of magnitude, allowing it to tackle the problem more efficiently. 
This has already allowed explicit computation of $\mu-T$ and $y-T$ distortion correlation signals for a number of scenarios \citep{kite_spectro-spatial_2023-III}, demonstration that valuable constraint can be derived with existing data from \Planck, and in the future {\it Litebird} and possibly the SKA \citep{Remazeilles2018muT, RRC2022, BF2022, Zegeye2024}.
Despite this success, there are still some discrepancies between the full {\tt CosmoTherm} solution for average SDs and the results obtained with this new FH framework that need to be overcome before starting to study the anisotropic case. 

In this work, many of the aforementioned aspects will be deepened. In particular, in Section \ref{sec:Greens} we will show the Green's functions and the relative temperature difference for various injection redshifts, extending the already known pre-recombination scenarios to late times. We will analyze the observed spectral features and the underlying physics, also investigating the effects of the Hubble cooling and details of the energy branching ratios. In Section \ref{sec:distortion_vis} we will show the distortion visibility function computed with {\tt CosmoTherm}, focusing specifically on low redshifts. And finally in Section \ref{sec:FH_treatment} we improve the FH treatment to reduce the current mismatch. To do so, we study the amplitude of the instantaneous energy injection and introduce a new effective critical frequency to improve the spectral evolution with the FH. This also reveals novel effects on the $\mu$-distortion shape that we capture by using a new spectral template computed with {\tt CosmoTherm}.

\begin{figure}
  \centering
\includegraphics[width=\columnwidth]{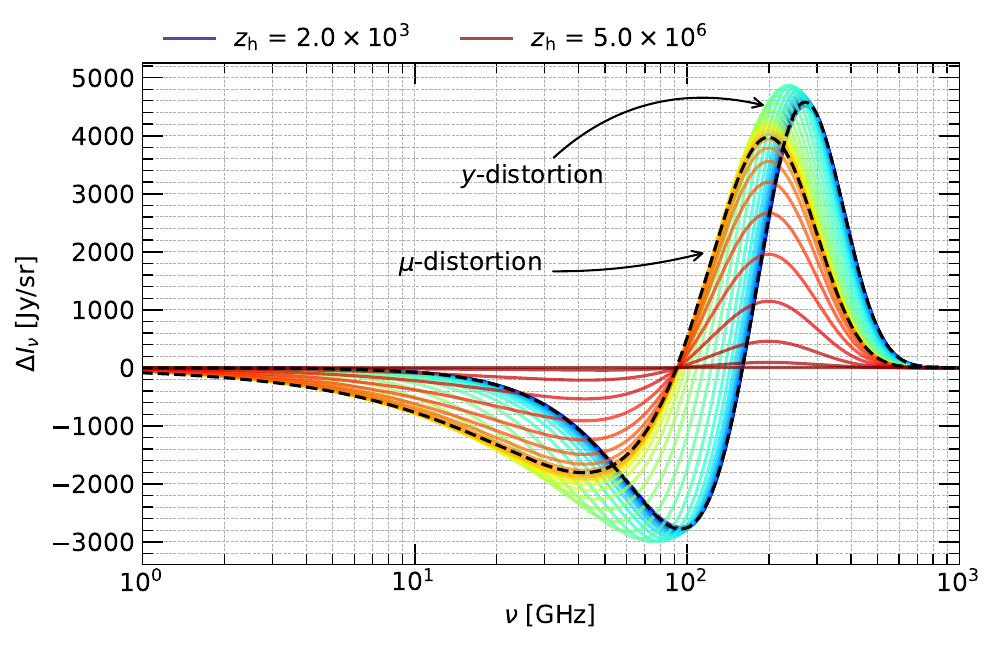}
  \\[1mm] \includegraphics[width=\columnwidth]{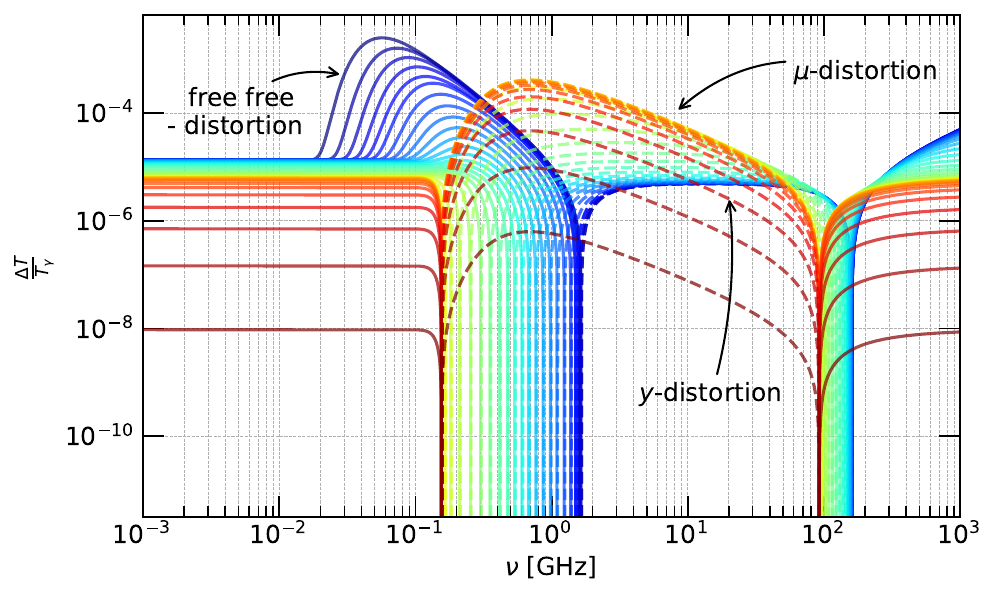}
  \\
    \caption{{\tt CosmoTherm} results for injections at $\zh\in(\pot{2}{3}, \pot{5}{6})$ and total energy release $\Delta \rho_\gamma/\rho_\gamma=10^{-5}$. In the upper panel, the variation of the Green's function according to the injection redshift can be observed. In the lower panel, the relative temperature difference is illustrated, highlighting the full range of frequencies and transition of the early-type $\mu$- to late-type $y$-distortions. Here, negative parts are shown as dashed lines, to better capture the large dynamic range of the signals.} 
\label{fig:earlyGF}
\end{figure}

\section{The distortion Green's function across cosmic time}
\label{sec:Greens}
In this section, we will focus on the computation of the Green's function, which contains fundamental information for the study of the thermalization process in our Universe. In particular, we will analyze its behaviour at different epochs covering a broad frequency range, highlighting the main features of the signal obtained. 

Relating the energy released in the early Universe by various processes to a distortion potentially observable today is generally hard. However, in the limit of small distortions and fixed background cosmology, computing the Green's function with {\tt CosmoTherm} can help simplify the problem \citep[e.g.,][]{Chluba2013Green}. In this framework, the spectral distortion intensity at $z=0$ can be expressed as follows:
\begin{equation} \label{eq::greens}
    \Delta I_{\nu} = \int G_{\rm th}(\zh,\nu) \frac{\id (Q/\rho_{\gamma})}{\id \zh} \id \zh\,,
\end{equation}
where $G_{\rm th}(\zh,\nu)$ is the Green's function that has to be computed numerically and $ {\id} (Q/\rho_{\gamma})/{\id} z_{\rm h}$ is the comoving relative energy injection rate. For an instantaneous energy release, envisioned as a $\delta$-function energy injection rate, the distortion $\Delta I_{\nu}$ reduces to the Green's function $G_{\rm th}(\zh,\nu)$ itself. In this limit, the Green's function at various redshifts can be obtained numerically using {\tt CosmoTherm} and then tabulated.

\subsection{The early-time distortion Green's function}
\label{sec:Greens_early}
In the upper panel of Fig.~\ref{fig:earlyGF}, we illustrate the distortion Green's function for several cases focusing on early times ($\zh\geq \pot{2}{3}$). We removed any temperature shift contribution [i.e., the full thermalization spectrum, $G(x)=x \, \expf{x}/(\expf{x}-1)^2$ to focus on the distortion signal only. Where $x=h\nu/k_{\rm B}\Tg$ is the with dimensionless frequency and $\Tg$ the reference blackbody temperature used in the computation\footnote{Usually $\Tg\approx \TCMB$ for all purposes.}. At $\zh\gtrsim \pot{3}{5}$, the Green's function has a spectrum that is extremely close to a $\mu$-distortion, with diminishing amplitude at $\zh\gtrsim \pot{2}{6}$ due to efficient thermalization \citep{Sunyaev1970mu, Burigana1991, Hu1993}. For $\zh\lesssim \pot{3}{5}$, one can observe a smooth transition from a $\mu$- to a $y$-distortion just before the recombination epoch, with non-$\mu$/non-$y$ contributions in the intermediate era ($10^4\leq \zh \leq 3 \times 10^5$), where the spectral shape is characterized by a superposition of $\mu$ and $y$, plus a residual distortion \citep{Chluba2013PCA}. The latter contains epoch-dependent information \citep{Illarionov1975, Hu1995PhD, Chluba2011therm, Khatri2012mix}, which can be used to distinguish various energy release scenarios \citep[e.g.,][]{Chluba2013fore, Chluba2013PCA}.

In the lower panel of Fig.~\ref{fig:earlyGF}, instead, we represent the distortion in terms of the effective brightness temperature variation, $$\frac{\Delta T}{T}= \frac{T(x) - \Tg}{\Tg},$$ where $T(x)$ is the effective temperature of the photon spectrum, obtained by comparing the distorted spectrum to a pure blackbody at each frequency. 
In the standard CMB bands ($\nu \simeq 10-1000\,\GHz$), we can observe the characteristic regimes of a $\mu$- and $y$-distortion. In contrast, at very low frequencies, the temperature difference {\it always} approaches a constant value $T(x) \approx \Te$ given by the electron temperature $\Te$ at the redshift considered, regardless of the high-frequency distortion. This can be understood as follows: we know that the emission and absorption processes are driven by the electrons, with increasing efficiency towards low frequencies, as suggested by the corresponding Boltzmann collision term for the photon occupation number $n$ \citep[e.g.,][]{Burigana1991, Hu1993, Chluba2018Varenna}:
\begin{equation}
\left. \frac{\partial n}{\partial \tau} \right |_{\rm em/abs} \approx \frac{K\expf{-\xe}}{x^3}[1-n\,(\expf{\xe}-1)]\,.
\end{equation}
Here, $\xe=h\nu/k\Te$ and $K$ is the double Compton (DC) and Bremsstrahlung (BR) emission coefficient that depends on frequency, the ionization state of the plasma and the electron and photon temperatures. 
As the free-free timescale decreases towards $x\rightarrow 0$, the photon distribution is pushed towards $n\rightarrow 1/(\expf{\xe}-1)$.
Therefore, electrons and photons are in equilibrium at the same temperature $T(x) \rightarrow \Te$ at very low frequencies, and the departures are exponentially suppressed. 

For the considered cases, we can also see a characteristic minimum of $\Delta T/T$ at $\nu\simeq 1\,\GHz$ during the $\mu$-era, which then morphs to a maximum located at $\nu \simeq 50-100\,\MHz$ in the $y$-era. This is related to the decrease of the Compton scattering efficiency at $\zh\lesssim \pot{3}{5}$ and the dominance of the free-free process at low frequencies \citep{Chluba2015GreensII, Cyr2024SPH}. The position of the low-frequency turnover again contains valuable epoch-dependent information about the energy-release process and thermal history.

\begin{figure}
  \centering
  \includegraphics[width=\columnwidth]{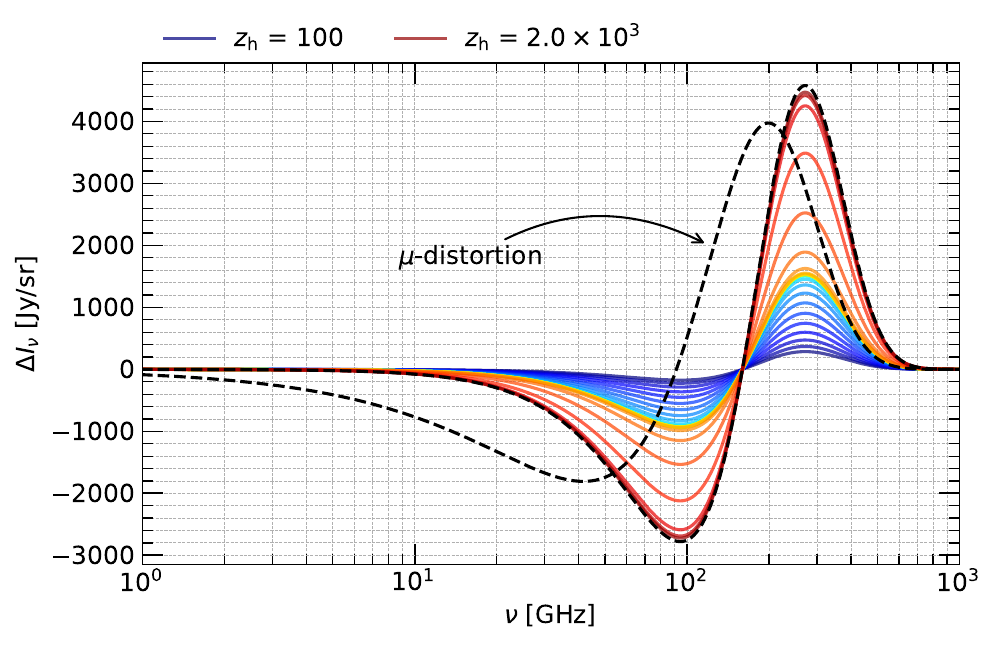}
  \\[1mm]
  \includegraphics[width=\columnwidth]{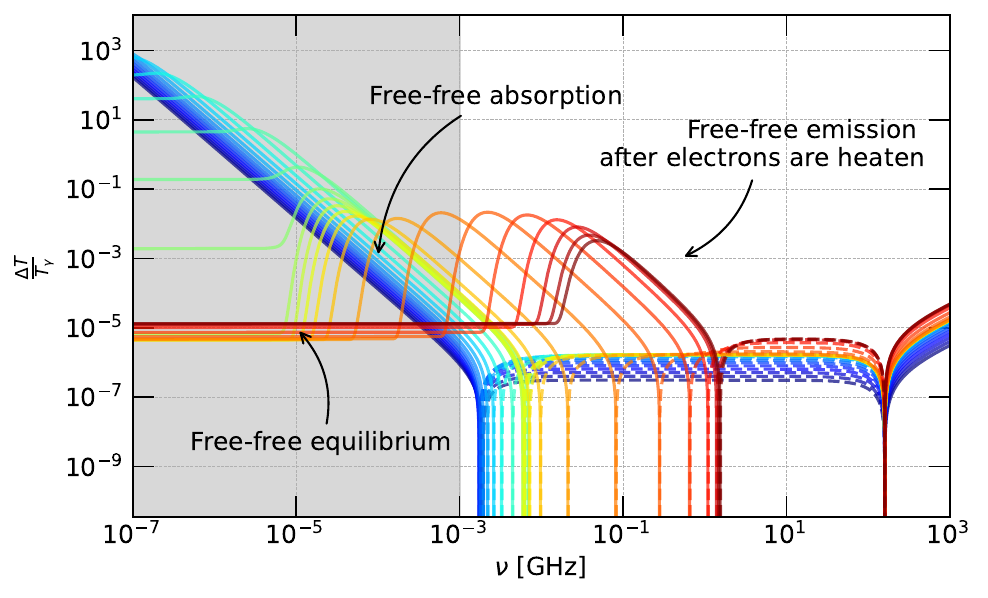}
  \\
  \caption{As in Fig.~\ref{fig:earlyGF} but for $\zh\in(10^2, \pot{2}{3})$. At high frequencies ($\nu\gtrsim 10\,\GHz$), the dominant spectral shape is $Y(x)$ with varying amplitude. The relative temperature difference instead, shows a frequency-dependent behaviour at low frequencies which in principle encodes valuable epoch-dependent information. The gray shaded area represents a frequency range that will not be reached by near future experiments.} 
  \label{fig:lateGF}
\end{figure}

\subsection{The late-time distortion Green's function}
\label{sec:Greens_late}
In contrast to previous works, here we are also interested in the late-time distortion Green's function ($\zh\lesssim 10^3$) reaching deep into the post-recombination era, as illustrated in Fig.~\ref{fig:lateGF}.
At high frequencies ($\nu\gtrsim 10\, \GHz$), the distortion is always close to a $y$-distortion, since Comptonization is rather inefficient at $z\lesssim 10^4$ and no $\mu$- or residual distortion contributions are created. However, one can notice that the amplitude of the distortion decreases with the injection redshift, although in the computations the amount of energy injected is kept the same. The explanation is the lower efficiency of electrons transferring their energy to the photons through Compton scattering. In fact, when some energy is injected into the medium, it first raises the electron temperature, which subsequently leads to up-scattering of the photons interacting with them, sourcing a pure $y$-distortion. However, this kind of signal does not have epoch-dependent information about $\zh$, being just the same spectral shape with rescaled amplitude. Subtleties about the efficiency of converting the released energy into electron heating are discussed in Sect.~\ref{sec:xc_effective}.

The temperature difference signal reveals a different story, especially at the low-frequency end of the CMB spectrum (lower panel of Fig.~\ref{fig:lateGF}). As stated above, at very low frequencies, the photon distribution is directly coupled to the electrons, but the frequency at which the coupling is tight depends on the redshift. When the released energy heats the electrons, $\Te$ increases and the electrons emit a free-free type spectrum, $\Delta T/T \simeq 1/x$. As time advances, photons are absorbed at very low frequencies and pushed towards equilibrium with the electrons. As the electrons slowly cool, a free-free self-absorption peak in $\Delta T/T$ is formed at $\nu \lesssim 100\MHz$. The position of the peak depends on the exact injection moment. Therefore, by measuring the low-frequency signal, one could in principle extract detailed information about the late-time thermal history. 

However, observationally, extracting this information will be quite challenging. Absolute measurements at $\nu\simeq 10-20\,\GHz$ are the target of TMS \citep{Jose2020TMS} in the next years. Below this, distortion measurement might also become possible with APSERa \citep{Mayuri2015, Krishna2024} and L-BASS \citep{ZerafaPhD2021}, in principle allowing to extract additional redshift-dependent information mentioned above. At radio frequencies, absolute measurements may become feasible with RHINO \citep{bull2024RHINO}, although reaching mK sensitivity poses a significant challenge. In addition, free-free emission from galaxies including our own \citep[see][for discussion of the main distortion foregrounds]{abitbol_pixie, Rotti2021SD} contributes a significant foreground \citep[also see discussion in][]{Trombetti2014ff}, although the presence of an abrupt self-absorption peak at $\nu \simeq 100\,\MHz$ is not expected. Nevertheless, we show the spectra as an illustration of the thermalization physics.

\begin{figure}
    \centering
    \includegraphics[width=1\linewidth]{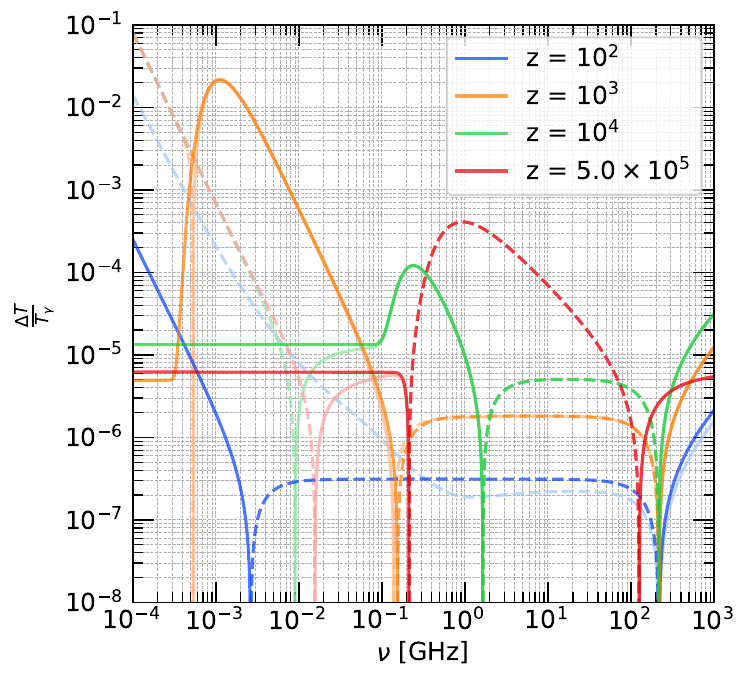}
    \caption{Relative temperature difference for instantaneous energy injections at three redshifts. Heavy lines are used for the scenario for which the Hubble cooling effect is not included, while light lines show the cases with Hubble cooling. The main difference is observed at low frequencies, where the $T(x)$ drops drastically.}
    \label{fig:DT-hc}
\end{figure}

\subsection{Effect of Hubble cooling}
\label{sec:hc}
In this section, we illustrate some additional details in the computation of the distortion Green's function. Specifically, we want to illustrate the effects that Hubble cooling has on the evolution of the distortion signal. This inevitable energy extraction from the CMB \citep{Chluba2005, Chluba2011therm, Khatri2011BE} was excluded above, although it modifies the signal at low frequencies.

The physics is as follows: without Compton interactions, electrons would cool as $\Te \propto (1+z)^{2}$. However, energy is transferred from the photons to the electrons while they are thermally coupled (i.e., down to $z\simeq 200$), forcing $\Te \simeq \Tg\propto (1+z)$. To maintain this equilibrium state, photons have to lose some of their energy, generating a negative $\mu\simeq - \pot{3}{-9}$ \citep{Chluba2005, Chluba2016}. 

In this context, when we say that Hubble cooling is not considered, we mean that the corresponding term in the energy exchange rates is switched off and the two temperatures scale the same at all redshifts. Including this effect does not alter the distortion shape significantly at high frequencies ($\nu\gtrsim 1-10\,\GHz$), since the related $\mu$/$y$ contributions are about four orders of magnitude below the considered energy release in our examples (i.e., $\Delta \rho_\gamma/\rho_\gamma=10^{-5}$). However, what changes remarkably is the effective brightness temperature at frequencies below $\simeq 50\,\MHz$, as shown in Fig.~\ref{fig:DT-hc}: we can observe that $\Delta T(x)/T$ plummets drastically at some low frequency that depends on the considered case. As we said before, with Hubble cooling $\Te$ decreases more rapidly than the photon temperature after thermal decoupling, leading to $\Te<\Tg$. This leads to net free-free {\it absorption} of photons at very low frequencies, where this process is most efficient, resulting in the observed solutions. 

The discussion above illustrates that the low-frequency CMB spectrum ($\nu\lesssim 50-100\,\MHz$) is highly sensitive to details of the electron/baryon temperature evolution. However, this brings us to another important aspect of the discussion. Once the first astrophysical sources appear and reionization sets in, the low-frequency spectrum will further evolve significantly at $z\lesssim 10$. The detailed distortion shape directly depends on the heating rates of the medium and also if photons are injected \citep{Chluba2015GreensII, Bolliet2020PI, Acharya2023SPH}. Due to the inefficiency of Comptonization, these effects can be captured primarily analytically, essentially including only the source of $y$-type distortions modified by the effects of free-free emission and absorption on the total CMB spectrum \citep{Cyr2024SPH}. The onset of reionization is also expected to cause significant anisotropies in the background spectrum \citep[e.g.,][]{Burigana2004ff, Trombetti2014ff}, which could be targeted using cross-correlation studies. We leave a more detailed consideration to future work.

\vspace{-3mm}
\section{Distortion visibility function}
\label{sec:distortion_vis}
We are now interested in how much energy is still visible today as a distortion after a single injection. This energy fraction is referred to as the distortion visibility function\footnote{In existing literature, $\Jdiss$ is also often denoted as $\Jbb$.}, $\Jdiss$, which depends on the injection redshift and can be obtained as an output from {\tt CosmoTherm} when computing the Green's function. Previous studies have been limited to the pre-recombination era ($\zh \geq 10^3$), and here we now also consider the post-recombination era. 

\begin{figure} 
  \centering
  \includegraphics[width=\columnwidth]{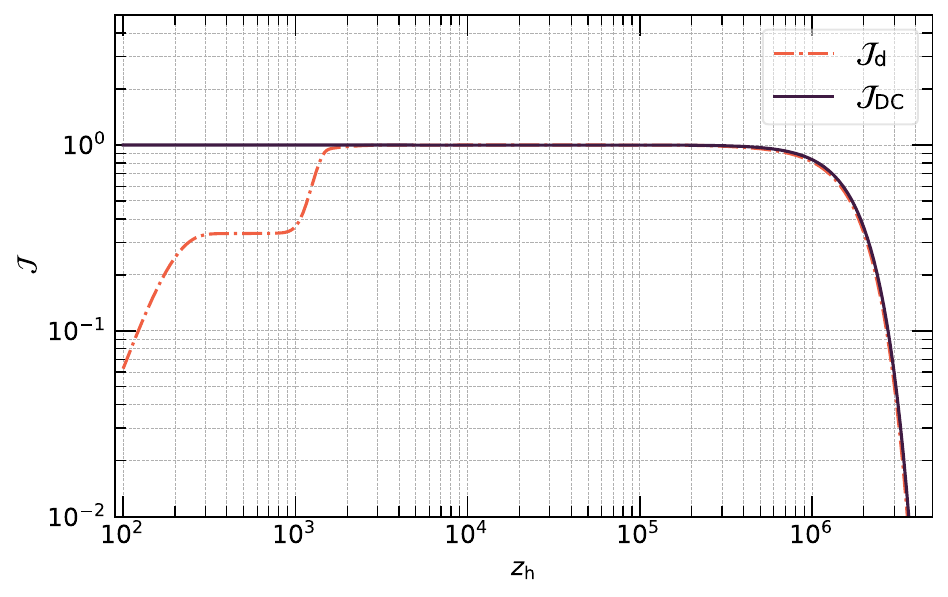}
  \\
  \includegraphics[width=\columnwidth]{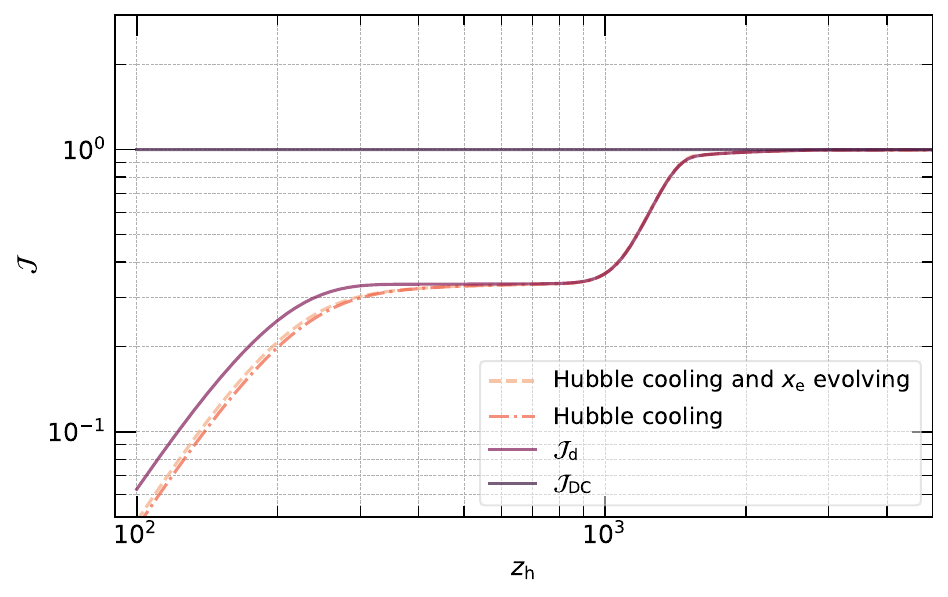}
  \\
  \caption{Distortion visibility function for a wide range of injection redshifts, also covering the post-recombination era. For reference, we also show the analytic approximation, $\Jdc$. The numerical results obtained with {\tt CosmoTherm} depart slightly at $\zh\geq 10^6$ and very visibly at $z\lesssim \pot{2}{3}$. In the lower panel, we focus on the low-redshift evolution and compare various cases as explained in the main text.}  
  \label{fig::Jbb}
\end{figure}

The distortion visibility function was analyzed in several studies \citep[e.g.,][]{Chluba2005,
Chluba2011therm, Khatri2012b, Chluba2015, Chluba2020large}. It also appears in analytic approximations of the Green's function \citep[cf.,][]{Chluba2013Green}:
\bsub
\begin{align}
    G_{\rm th}(z_{\rm h}, \nu) 
    &\approx  1.401 \mathcal{J}_{\mu}(\zh)\,M(x)
    + \frac{\mathcal{J}_y(z_{\rm h})}{4}\,Y(x) + \frac{\mathcal{J}_{T}(\zh)}{4}\,G(x)\,,
\\
M(x)&=G(x)\left[\frac{1}{\beta_M}-\frac{1}{x}\right], 
\;\; Y(x)=G(x)\left[x\,\frac{\expf{x}+1}{\expf{x}-1}-4\right]\,,
\end{align}
\esub
where $G(x)=x \expf{x}/(\expf{x}-1)^2$ is the spectrum of a small temperature shift, and $M(x)$ and $Y(x)$ are the standard $\mu$ and $y$-type distortion spectra with $\beta_M=2.1923$. Here we also introduced the $\mu$, $y$ and temperature visibility functions (i.e., the specific energy branching ratios), $\mathcal{J}_{\mu, y, T}$ which allow modelling of approximate proportions of these spectra to the total distortion signal.

Comparing Eq.~\eqref{eq::greens} with $\Delta I = \mu \, \Delta I_{M} + y \, \Delta I_{Y} + \Theta \, \Delta I_{G}$, where $\Theta$ determines the temperature shift, we then have
\begin{equation}
    s = \alpha_f \int^{\infty}_0 \mathcal{J}_s(z) \frac{{\id} (Q/\rho_{\gamma})}{{\id} z_{\rm h}} {\id}z\,,
\end{equation}
where $s\in \{\mu,y, \Theta\}$ is the amplitude of the distortion $f\in\{M, Y, G\}$ and $\alpha_f \in \{1.401, 1/4,1/4\}$ are energy normalization factors.

The distortion visibility function directly enters the definitions of $\mathcal{J}_s$. Each of them has a different redshift dependence and parametrizes the transition from one distortion to another. One common approximation, valid at $\zh\gtrsim 10^3$, is \citep[see][for comparison of various alternatives]{Chluba2016}
\bsub
\begin{align}
\mathcal{J}_y(\zh) & \approx \left(1+\left[\frac{1+\zh}{6.0\times 10^4}\right]^{2.58}\right)\,, 
\\
\mathcal{J}_{\mu}(\zh) &  \approx \left[1 - \exp\left( - \left[\frac{1+\zh}{5.8 \times 10^4}\right]^{1.88} \right)\right] \Jdiss(\zh)\,,
\\
\mathcal{J}_T(\zh)&\approx 1-\Jdiss(\zh)\,.
\end{align}
\esub
Analytically, $\Jdiss(\zh)\approx \Jdc(\zh)$, with the function
\begin{equation}
    \Jdc = \exp\left(-\left[\frac{\zh}{1.98\times 10^6}\right]^{5/2}\right)\,.
\end{equation}
This approximation neglects the effect of Bremstrahlung and various time-dependent corrections as well as relativistic corrections \citep[see][for detailed derivations]{Chluba2015}.

In Fig.~\ref{fig::Jbb}, we illustrate the distortion visibility function for several settings in {\tt CosmoTherm}. For comparison, we also show the analytic approximation $\mathcal{J}_{\rm DC}$. At $\zh>10^6$, $\mathcal{J}_{\rm DC}$ slightly overestimates the distortion visibility \citep{Chluba2005, Chluba2011therm, Khatri2012b, Chluba2014}, an aspect we will return to in Sect.~\ref{sec:FH_treatment}.  
However, more drastic differences are visible for $\zh < \pot{2}{3}$, where the curves diverge significantly. This behavior is related to the thermal decoupling of electrons and photons and also to the modeling of the heating efficiency, which for default settings follows \citet{Chen2004} (see Sect.~\ref{sec:xc_effective}). Broadly, when we inject energy into the medium, the electron temperature rises, which in turn leads to the transfer of energy to photons. If the two start to decouple, this process is less inefficient, and a smaller fraction of the injected energy reaches the photon field, as already mentioned in Sect.~\ref{sec:Greens_late}. 

The precise shape of the post-recombination distortion visibility depends on various settings. The lower panel of Fig.~\ref{fig::Jbb} highlights how the Hubble cooling accelerates the decoupling of photons and electrons at $\zh \lesssim 200$. The timescale for energy transfer from the electrons to the photons is \citep[e.g.,][]{Hu1995PhD, Chluba2018Varenna}
\begin{equation}
    t_{\rm e \gamma} \approx  \left[\frac{4 k \Te}{\me c^2} \,  \sigma_{\rm T} \Ne c \right]^{-1}\,,
\end{equation}
where $\theta_{\rm e} = 4 k \Te / \me c^2$ is the fraction of energy exchanged in each interaction and $t_{\rm T} = (\sigma_{\rm T} \Ne c)^{-1}$ is the Thomson scattering timescale. Therefore, a lower electron temperature results in a longer timescale and reduced energy transfer, as seen in the figure. 

If we simultaneously evolve the free-electron fraction $\Xe$ and the distortion, the net effect is a slightly higher free-electron fraction. This is simply because hotter electrons, as expected after the injected energy heats them, recombine slightly less efficiently. This implies that $\Ne$ is larger and therefore $t_{\rm e \gamma}$ decreases. It leads to a tighter coupling and a small increase of the distortion visibility function (see Fig.~\ref{fig::Jbb}). This effect is much more subtle and depends on the exact recombination treatment, as we illustrate next. 

\begin{figure}
    \centering
    \includegraphics[width=1\linewidth]{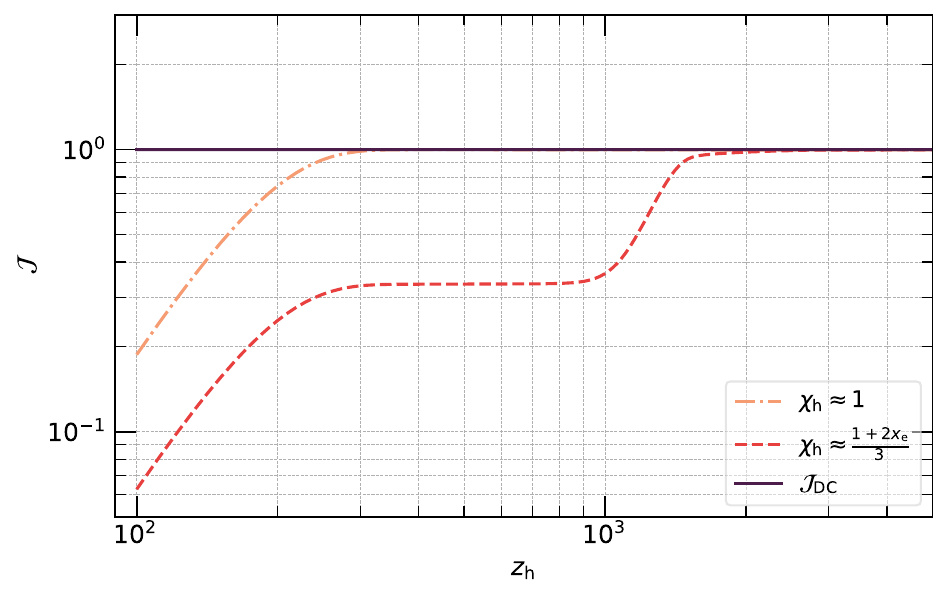}
    \caption{Low-redshift distortion visibility functions with different branching ratios for heating. The heating functions affect both the $\Te$ and $\Ne$ evolution, requiring a combined treatment.}
    \label{fig:Jbb-branching}
\end{figure}

\subsection{The effects of energy branching ratio}
\label{sec:gheat}
Another aspect that should be considered is that not all the energy released heats the electrons of the medium, which in turn transfer their energy to the photons, generating a spectral distortion. 
In high-energy particle cascades, a fraction of the energy also goes into ionization and excitation of atoms \citep{Chen2004, Huetsi2009, Slatyer2009, Chluba2010a}. Following \citep{Chen2004}, the approximate branching ratio is:
\begin{align}
\chi_{\rm h} &\approx \frac{1+2 \Xe}{3}\,, \qquad \chi_{\rm e} \approx \chi_{\rm i} \approx \frac{1- \Xe}{3}\,,
\end{align}
for heating, excitation, and ionization, respectively.
In the fully ionized medium, all energy goes into heating the electrons. This can temporarily cause non-thermal relativistic electrons, which modify the distortion shape \citep{Slatyer2015, Acharya2019a}; however, this is not modeled here.

For default settings, {\tt CosmoTherm} uses the above branching ratios. However, in the study of the Green's functions, we are only interested in the energy that reaches the CMB. It is then reasonable to modify the code so that all the energy injected causes heating of the electrons, separating the effects. This treatment becomes exact when the distortion is created only by heating processes without any extra ionizations or excitations that affect the recombination process.

In Fig.~\ref{fig:Jbb-branching}, we show the comparison between the default and modified $\chi_{\rm h}=1$ treatments. Evidently, the distortion visibility for pure heating remains close to unity (i.e., all the energy release causes a $y$-distortion) until significantly lower redshifts than for the default setting. Our results illustrate the range of responses that one could expect. The respective energy branching ratio can be directly captured by introducing an effective heating rate, i.e., $\dot Q^*=\chi_{\rm h}\dot Q$.

\section{Improved frequency hierarchy treatment}
\label{sec:FH_treatment}
Instead of solving the full thermalization problem, it was recently realized that the problem can be simplified using a special spectral basis \citep{chluba_spectro-spatial_2023-I}. However, this came with some shortcomings, and in this last section, we want to study whether they can be overcome through a better matching with {\tt CosmoTherm}.

\begin{figure*}
\centering
  \includegraphics[width=\columnwidth]{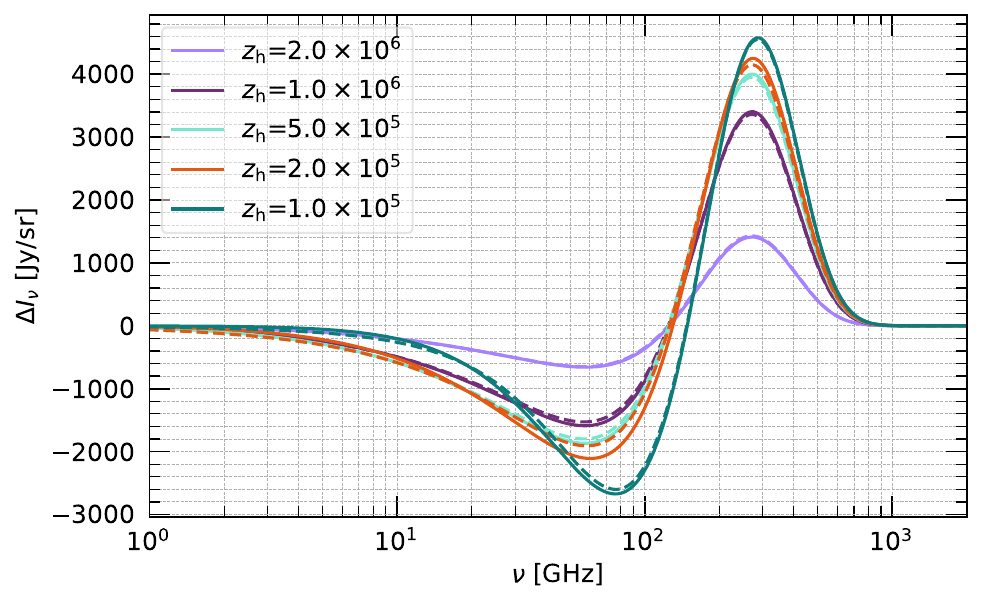}
  \hspace{2mm}
  \includegraphics[width=\columnwidth]{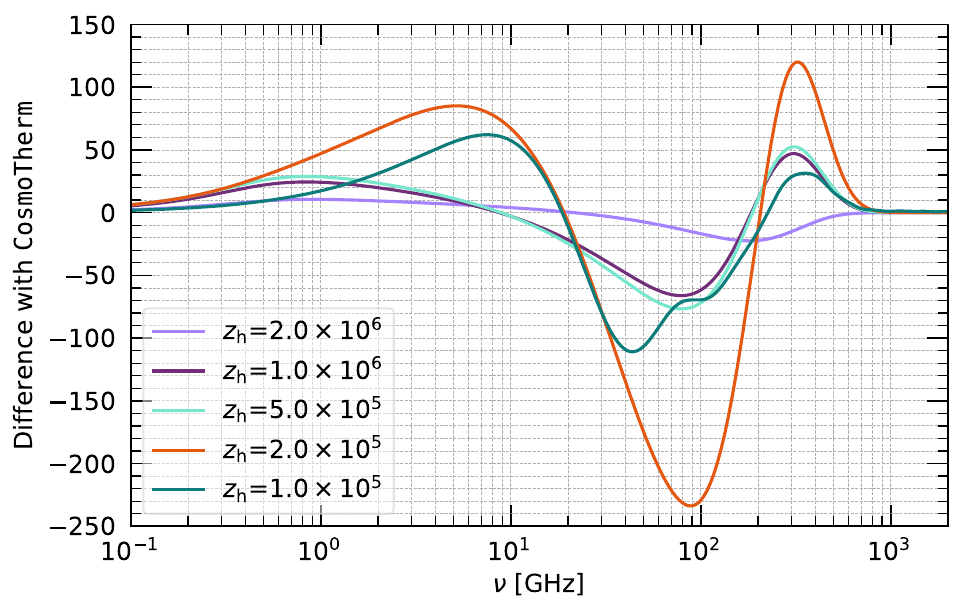}
  \\
  \caption{Comparison of the full {\tt CosmoTherm} results (solid lines) with the approximate FH treatment (dashed lines) for single injections at various heating redshifts and $\Delta \rho_\gamma/\rho_\gamma=10^{-5}$. For the FH we used a spectral basis up to $Y_{15}$.
    The left panel illustrates the distortion and the right panel shows the difference between the two treatments.
    The mismatch increases with decreasing $\zh$ and generally has a relative level of $\simeq 10\%-20\%$.
    }
    \label{fig:FH_cases}
\end{figure*}
\subsection{Basics of the frequency hierarchy treatment}
\label{sec:FH_treatment_basics}
The evolution of the photon occupation number distortion, $\Delta n(t,x) = n - n_{\rm bb}$, with respect to the blackbody distribution, $n_{\rm bb}$, is described by a kinetic equation, which takes into account Compton scattering, emission/absorption processes and external heating. This complicated thermalization problem can be cast into a much simpler matrix equation \citep{chluba_spectro-spatial_2023-I}, which then also opens possibilities for studying spectral distortion anisotropies \citep{chluba_spectro-spatial_2023-II, kite_spectro-spatial_2023-III}. Defining the distortion vector $\bm{y} = (\Theta,y,y_1,...,y_{\rm N}, \mu)^{\rm T}$, the decomposition used for the computation is the following:
\bsub
\begin{align}
    &\Delta n (t,x) = \bm{B}(x) \cdot \bm{y}(t)\,, \\
    & \bm{B} = (G(x), Y(x) , Y_1(x), ..., Y_{\rm N} ,M(x))^{\rm T}\,, 
\end{align}
\esub
with $Y$, $M$, and $G$ being the standard distortion shapes, and where $Y_k$ is obtained using the boost operator, ${\bm\hat{\mathcal{O}}}_x=-x\partial_x$ on the $y$-distortion, $Y_{k} = (1/4)^k {\bm\hat{\mathcal{O}}}^k_x Y$. With this \textit{Ansatz} one finally obtains the system:
\bsub
\label{eq:FH_system}
\begin{align}
    &\dot{{\bm y}} \approx \Dot{\tau} \,\Thg \left[ \bm{M}_{\rm K} \, \bm{y} + \bm{D} \right] + \frac{\Dot{\bm{Q}}}{4}\,,\\
    &\bm{D} = ( \gamma_T \xc \mu, 0,0,...,0, -\gamma_N \xc\, \mu)^{\rm T}\,,\\
    &\Dot{\bm{Q}} = (0,\Dot{Q}/\rho_\gamma, 0,...,0,0)^{\rm T}\,,
\end{align}
\esub
where $\bm{M}_{\rm K}$ is the Kompaneets mixing matrix of the basis, describing how the distortion transforms in time. The term $\bm{D}$ describes how quickly $\mu$ is converted into a temperature shift, and $\dot{\bm{Q}}$ is the heating source term, which only causes a $y$-distortion. We also introduced the critical frequency for photon emission, $\xc$, which we discuss in Sect.~\ref{sec:xc_effective}. The emission coefficients are given by $\gamma_N \approx 0.7769$ and $\gamma_T \approx 0.1387$, and the Thomson optical depth $\tau = \int \sigma_{\rm T} N_{\rm e}\, c \id t$. The 'dot' denotes the derivative with respect to time. Henceforth, we refer to Eq.~\eqref{eq:FH_system} as the {\it frequency hierarchy treatment} (FH treatment).

The benefit of the FH treatment given above is that it significantly accelerates the computation. The results can be compared to the exact result of {\tt CosmoTherm}, which instead solves thousands of equations. For this purpose, one can consider a single injection scenario using a narrow Gaussian with a peak at $\zh$ and truncate the hierarchy at $N_{\rm max} = 15$. The compatibility has already been considered \citep[e.g., Sect.~2.8 of][]{chluba_spectro-spatial_2023-I}, and here we will focus in particular on the early injections at $\zh\geq 10^5$. 

In Fig.~\ref{fig:FH_cases}, we illustrate the departures for several injection redshifts. Differences are generally more noticeable at $\nu\lesssim 100\,\GHz$. It is also clear that at $\zh\simeq \pot{2}{5}$ the mismatch is the largest relatively speaking. As stressed in \citet{chluba_spectro-spatial_2023-I}, additional spectral contributions that cannot be captured by the FH become relevant there, and achieving improvements will likely require extensions of the basis. However, at $\zh\gtrsim \pot{3}{5}$ one expects the distortion to be primarily represented as a $\mu$-distortion. Still, a mismatch is found. Possible causes are related to
the setup of the single injection scenarios and the approximations for the thermalization efficiency parametrized by $\xc$, which we investigate now. 

\subsection{Treatment of the single injection scenario}
\label{sec:inst_inj}
In the solutions presented above, it is assumed that the energy is injected instantaneously as a $\delta$ function. Computationally, this is approximated by a narrow Gaussian function, whose relative width has to be chosen carefully. The width of a Gaussian depends both on the mean value (i.e. injection redshift) and the relative standard deviation, and it is formally defined as $\Delta z = \zh\, \sigma$, so the energy release history is given by
\begin{equation}
     f(z,\zh,\sigma)= \frac{f_0}{\sqrt{2 \pi}\,\zh\sigma}\expf{-\frac{(z/\zh-1)^2}{2 \sigma^2}}\,,
\end{equation}
where $f_0$ determines the total energy release.
For the FH treatment, we usually use $\sigma=0.01$; however, this is folded with the sharply decreasing distortion visibility function, which could explain the mismatch seen early on. 
Therefore, we now quantify how the final result is influenced by the value of $\sigma$. This will help to choose the best value to compare the results obtained using the frequency hierarchy setup with the {\tt CosmoTherm} solution.

At high redshifts, the distorted spectrum is roughly given by
\begin{equation}
\Delta I \approx \Delta I_M \int \Jdiss(z) \, f(z,\zh,\sigma) \id z\,,
\end{equation}
where we substituted the Green's function in the $\mu$-epoch. This result is proportional to the average of the distortion visibility function computed over a Gaussian function:
\begin{equation}
    \langle \Jdiss \rangle = \int \Jdiss(z) \, \hat{f}(z,\zh,\sigma) \id z\,,
\end{equation}
where $\hat{f}(z,\zh,\sigma)=\hat{f}(z,\zh,\sigma)/f_0$. If $\hat{f}$ were a $\delta$ function, we would exactly recover $\Jdiss$. To quantify the effect of $\sigma$, it is therefore reasonable to compare the latter with its average value, $\langle \Jdiss \rangle$, at different injection redshifts and for various values of $\sigma$. 

\begin{figure}
    \centering
    \includegraphics[width=\columnwidth]{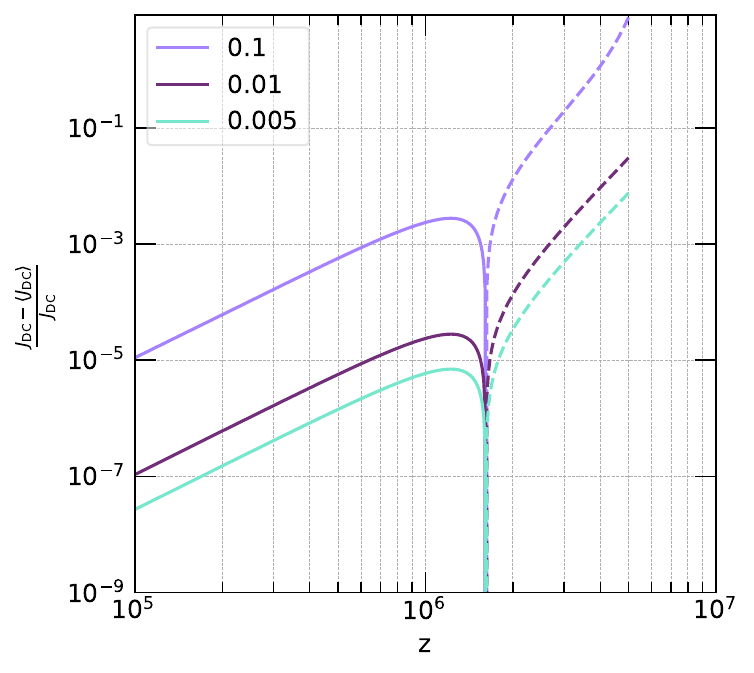}
    \\
    \caption{Illustrations of how well a Gaussian energy injection history approximates a Dirac-$\delta$ as a function of redshift and for various values of the relative width. At $z\lesssim 10^5$, the mismatch is negligible for all cases, while the precision can be hampered significantly early on.}
    \label{fig:rw}
\end{figure}

Fig.~\ref{fig:rw} shows the results computed with {\tt Mathematica}. To simplify the problem, we assumed that $\Jdiss\approx\Jdc$; however, this should give qualitatively similar results.
It is clear that the Gaussian is not a good approximation of a Dirac-$\delta$ function both for a large $\sigma$ and at high redshifts. The last aspect is related to the shape of $\Jdiss$, which decreases exponentially for redshift $\zh>10^6$. However, for $\sigma\lesssim 10^{-2}$ the discrepancy remains small compared to the order of magnitude seen in Fig.~\ref{fig:FH_cases}, ruling out this choice as the cause of the mismatch. In {\tt CosmoTherm}, $\sigma=0.001$ is used in the Green's function setup.

\begin{figure}
    \centering
\includegraphics[width=\columnwidth]{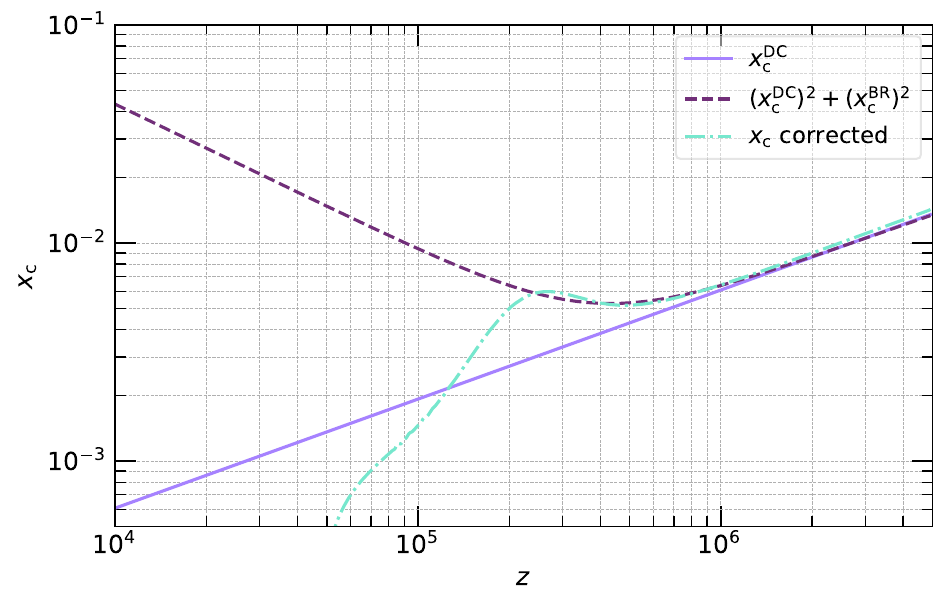}
\\
    \caption{Critical frequencies in function of redshift compared to the effective critical frequency corrected with the numerical values of $\Imu$. The main difference is observed for $z \lesssim 10^6$.}
    \label{fig:xc}
\end{figure}

\subsection{Effective critical frequency and \texorpdfstring{$\Imu$}{Imu} estimation}
\label{sec:xc_effective}
A parameter that has a significant impact on the shape of the spectral distortion is the critical frequency $\xc$. It is defined as the frequency where Compton scattering balances the emission/absorption processes, and it depends on redshift. An expression can be determined separately for DC or BR alone and then combined to obtain the total critical frequency $\xc^2 \approx (\xc^{\rm DC})^2+ (\xc^{\rm BR})^2$. For the DC case, the solution can be found analytically, using the evolution equations and assuming a stationary state \citep{Chluba2014}:
\bsub
\begin{align}
    \xc^{\rm DC,0} &\approx \sqrt{K_{\rm DC}/\theta_{\gamma}} \approx 8.60 \times 10^{-3}\left[\frac{1+z}{2 \times10^6}\right]^{1/2}\,,\\
    \xc^{\rm DC} &\approx \xc^{\rm DC,0} \left( 1+\frac{1}{2}\xc^{\rm DC,0}\right)^{1/2}  \left(1+14.16 \theta_{\gamma} \right)^{-1/2}\,,
\end{align}
\esub
where $\Thg=\kB\Tg/\me c^2$, and $\xc^{\rm DC,0}$ does not include relativistic corrections, while $\xc^{\rm DC}$ includes leading-order terms. 

For BR, the problem has been solved numerically, assuming the helium mass fraction to be $Y_{\rm p} = 0.24$:
\begin{equation}
    \xc^{\rm BR} \approx 1.23 \times 10^{-3}\left[\frac{1+z}{2 \times10^6}\right]^{-0.672}\,.
\end{equation}
This contribution is dominant at low redshifts but can be neglected instead in the early Universe, as also shown in Fig.~\ref{fig:xc}.

\begin{figure*}
\centering
\includegraphics[width=0.98\columnwidth]{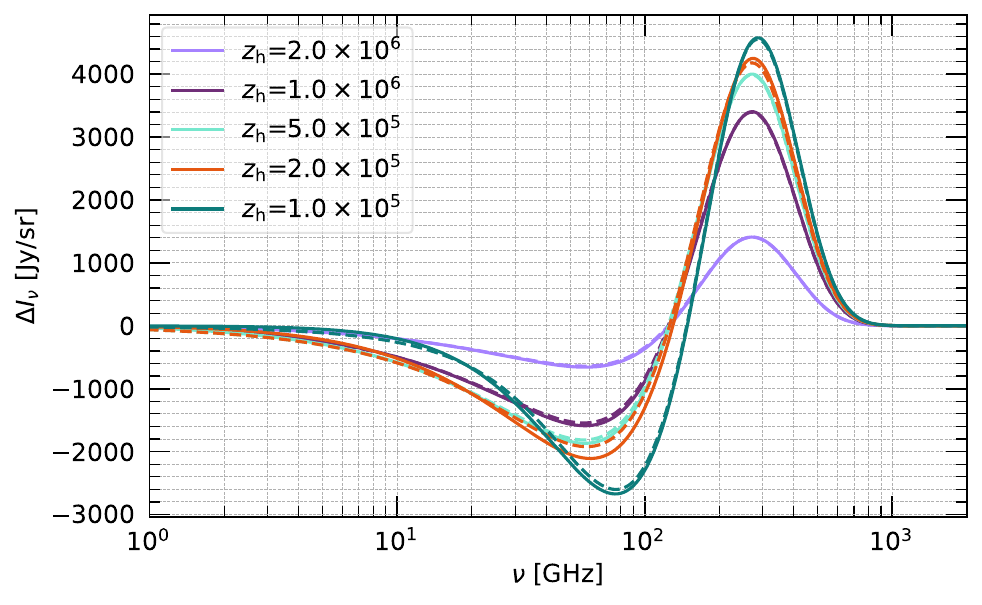}
\hspace{2mm}
\includegraphics[width=0.98\columnwidth]{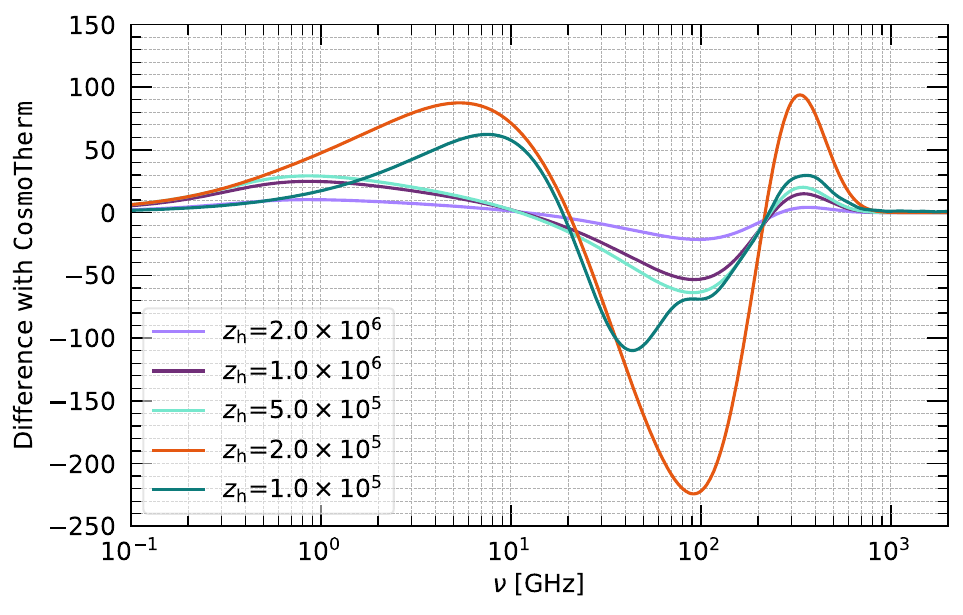}
\\
\includegraphics[width=0.98\columnwidth]{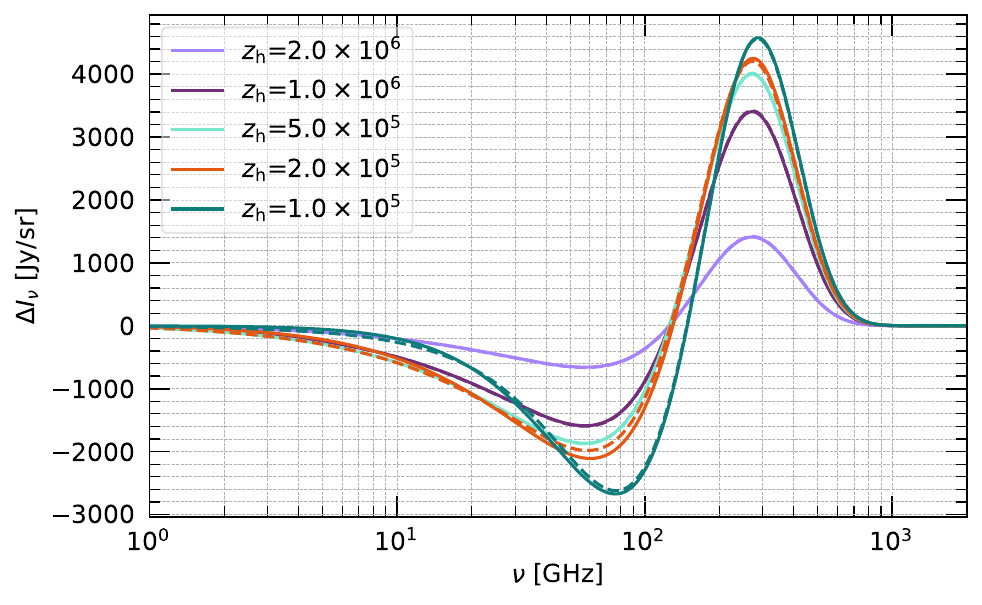}
\hspace{2mm}
\includegraphics[width=0.98\columnwidth]{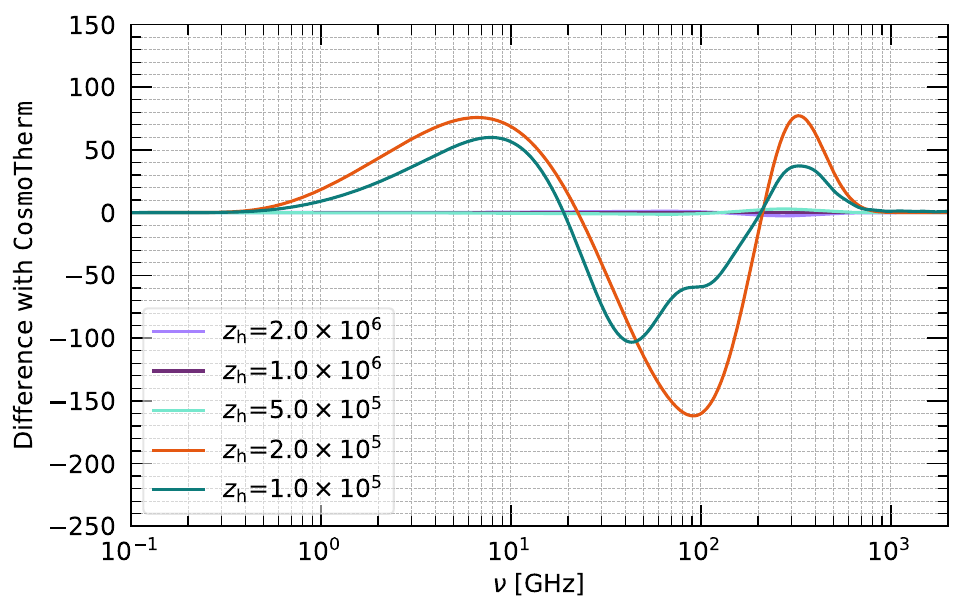}
\\
  \caption{Spectral distortion signals as in Fig.~\ref{fig:fhm-5e5-final} but the spectral basis is improved as discussed in Sect.~\ref{sec:FH_treatment}. In particular, the upper panels show what happens considering a value for $\Imu$ different from unity, and the lower panels show the results when also considering the new numerical $M^*(x)$ spectral shape. The latter case presents a significant improvement for $\zh\gtrsim \pot{3}{5}$, at both low and high frequencies.}
    \label{fig:fhm-5e5-final}
\end{figure*}
The critical frequency is particularly important because it appears in the equation for the evolution of the amplitude of the effective chemical potential $\mu(\tau,x) = \mu_{\infty}(\tau)\,\hat{\mu}(\tau,x)$, where $\mu_{\infty}(\tau)$ is the photon chemical potential at $x\gg 1$ \citep[see][for all details]{Chluba2014}:
\begin{equation} 
\label{eq:mu-evolution}
    \frac{\id \mu_{\infty}}{\id \tau} = 1.401\, \frac{\Dot{Q}}{\rho_{\gamma}}- \gamma_N x_c \theta_{\gamma} \Imu \, \mu_{\infty}\,.
\end{equation}
Here, $\Imu$ is a correction to the value of $\xc$ that depends on the details of the dynamics. In the FH treatment, $\Imu$ is set to unity. We now consider the corrections, which may lead to a better agreement with the full solution. 

To determine the best values for $\Imu$, we assume a single energy release, which allows us to neglect the source term in Eq.~\eqref{eq:mu-evolution} and find the solution:  
\begin{equation} 
    \mu_{\infty}(z) \approx 1.401 \Drr \,\mathcal{J}(\zh,z)\,,
\end{equation}
where $\mathcal{J}(\zh,z) = {\rm e}^{-\tau_{\mu}(\zh,z)}$ is the distortion visibility function computed between two different redshifts. Here, $$\tau_\mu(\zh,z)=\int_{z}^{\zh} \gamma_N \xc \Thg \Imu \, \frac{\sigT \Ne c }{H(z)(1+z)}\id z', $$ where we used $\id \tau/\id z=-\frac{\sigT \Ne c }{H(z)(1+z)}$ with the Hubble factor $H(z)$. If we consider the DC-only scenario, $\mathcal{J}(\zh,z)$ reduces to $\Jdc$ defined above. However, this also means that we can compute the values for $\Imu$ from the numerical result for $\Jdiss$. It is easy to show that $$\frac{\id \ln \Jdc}{\id z}\equiv \gamma_N \xc^{\rm DC,0} \Thg\,\frac{\sigT \Ne c }{H(z)(1+z)}=- \gamma_N \xc^{\rm DC,0} \Thg\,\frac{\id \tau}{\id z}.$$
Comparing with Eq.~\eqref{eq:mu-evolution}, one can find a definition for $\Imu$: 
\begin{equation} \label{eq:Imu}
    \Imu \approx \frac{\frac{\id \ln\Jdiss}{\id z}}{\frac{\id \ln \Jdc}{\id z}}= -\frac{\id z}{\id \tau} \frac{\id \ln\Jdiss}{\id z} \cdot \frac{1}{ \gamma_N \xc^{\rm DC,0} \theta_{\gamma}}\,.
\end{equation}
The numerical value for $\Imu$, including higher-order corrections, can then be computed using {\tt CosmoTherm} outputs. For the FH treatment, an improved effective critical frequency $\xc^{\rm eff}$ can now be defined as
\begin{equation} 
\xc^{\rm eff} = \xc^{\rm DC, 0} \cdot \Imu\,.
\end{equation}
The result for $\xc^{\rm eff}$ is also shown in Fig.~\ref{fig:xc}. At high redshifts, it is close to $\xc^{\rm DC, 0}$, with a nearly constant ratio of $\Imu\approx 1.0555$, which we use to extrapolate to the highest redshifts. This behavior indicates that the analytical and numerical curves scale the same at early times from Eq.~\eqref{eq:Imu}. The difference observed in Fig.~\ref{fig::Jbb} is then only due to a renormalization factor. Down to $z\simeq \pot{3}{4}$, $\xc^{\rm eff}$ closely follows the DC + BR value and then drops significantly at $z\lesssim \pot{2}{5}$. It was already noted in \citet{Chluba2014} that at these late times, the quasi-stationary approximation for the low-frequency evolution and interplay between photon emission and Comptonization breaks down such that a critical frequency treatment is no longer possible. Specifically, the transport of photons from $x\lesssim 10^{-2}$ to $x>1$ becomes extremely inefficient, even if, at $x\lesssim 10^{-2}$, significant spectral evolution is still found due to free-free emission (see Fig.~\ref{fig:earlyGF}).
However, at $z\lesssim 10^5$, not much conversion of $\mu\rightarrow T$ occurs, such that we simply use the numerical result for $\Imu$ down to $z = 4.5 \times 10^4$. Below this redshift, the critical frequency is set to zero in the FH treatment, with the understanding that the distortion evolution is then only modeled through the Kompaneets matrix.

The upper panels of Fig.~\ref{fig:fhm-5e5-final} show the distortion results obtained using the effective critical frequency $\xc^{\rm eff}$ in the FH treatment. At high frequencies, we can see a clear improvement of the match with the {\tt CosmoTherm} result, in particular for $\zh>\pot{2}{5}$, where the difference drops below $\simeq 1\%$. This is because a better timing for the conversion of $\mu \rightarrow T$ is obtained. 

\begin{figure}
    \centering
    \includegraphics[width=\columnwidth]{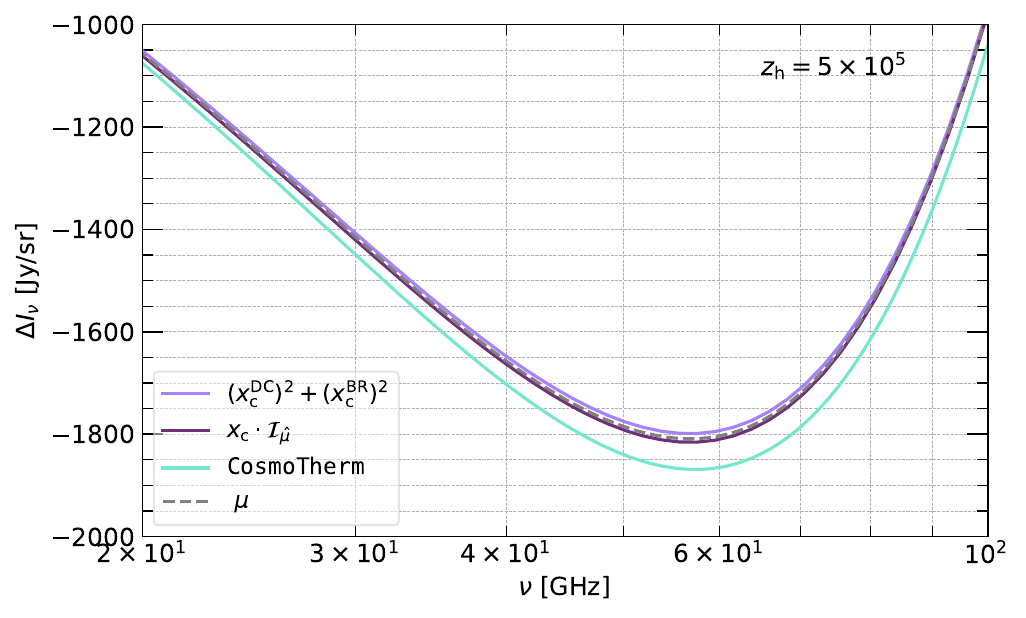}
    \caption{The low-frequency spectral distortion for various choices of $\xc$ in the FH treatment. It shows how using the precise value for $\Imu$ in the code influences the approximate solution, but not solving the problem.}
    \label{fig:fhm-5e5-corr}
\end{figure}
However, at low frequencies ($\nu\lesssim 100\,\GHz$), the mismatch remains unchanged. 
This is highlighted in Fig.~\ref{fig:fhm-5e5-corr} for injection at $\zh=5\times 10^5$. Although a small change is visible for various choices of the $\xc$ treatment, it is not enough to explain the disagreement with the full solution. Additional frequency-dependent corrections appear to be present, as also anticipated from analytic treatments \citep{Chluba2014}. However, a significant improvement has already been made in capturing the dynamics in the high-frequency part, which gives the most important contribution to the energetics.

\begin{figure}
    \centering
 \includegraphics[width=0.98\columnwidth]{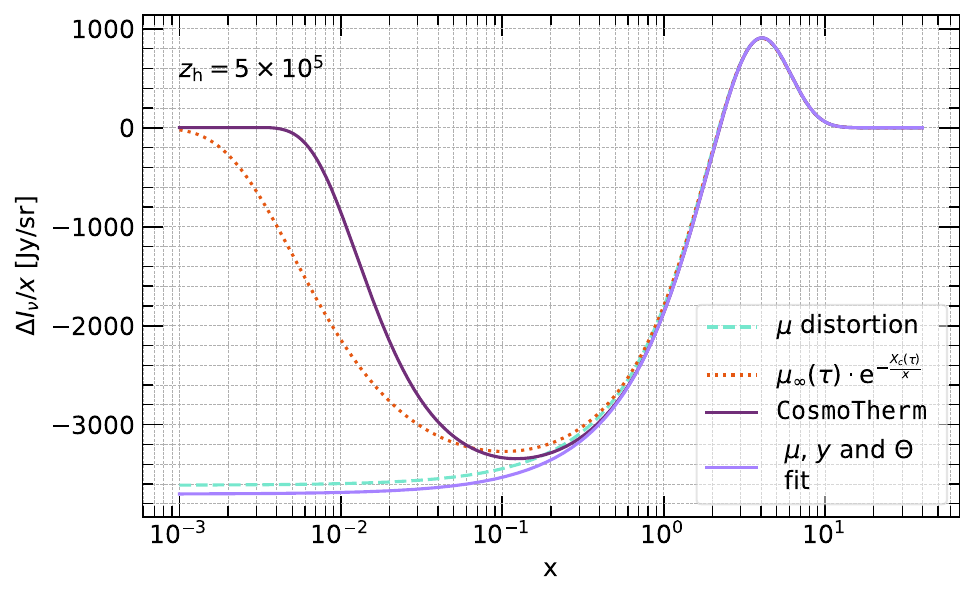}
    \caption{Distortion for a single $\Delta \rho_\gamma/\rho_\gamma = 10^{-5}$ divided by $x$ to highlight the low-frequency difference between the (number-conserving) {\tt CosmoTherm} solution, the analytical $\mu$-distortion and a direct fit using $\mu$, $y$ and $\Theta$. We furthermore also show the analytical $\mu$-distortion with the frequency-dependent amplitude to account for photon production effects at low frequencies.}
    \label{fig:fit}
\end{figure}
\subsection{The Numerical \texorpdfstring{$\mu$}{mu}-distortion template}
\label{sec:M_num}
In this section, we explore how the numerical solution can be better represented at high redshifts. For this, let us focus on the case for $\zh=\pot{5}{5}$ and $\Delta \rho_\gamma/\rho_\gamma = 10^{-5}$. In Fig.~\ref{fig:fit}, we show the {\tt CosmoTherm} result compared to various approximations. To emphasize the low-frequency part, we plot $\Delta I_\nu/x$, which in essence represents the specific photon number distortion, $\Delta N_\nu$.

At high frequencies ($x \gtrsim 2$), the {\tt CosmoTherm} solution is very close to a $\mu$-distortion with an amplitude that is roughly given by the analytic estimate,
$\mu_\infty \approx 1.401 \Jdc \, \Drr \approx 1.36 \times 10^{-5}$.
However, at intermediate ($x\simeq 0.1-1$) and low ($x\lesssim 0.1$) frequencies, significant deviations are observed from the simple distortion spectrum. Specifically, the numerical solution rapidly vanishes at low frequencies, an aspect that is not captured by the analytic approximation in terms of $M(x)$, which behaves as
$M(x) \xrightarrow{x \ll 1}{} -1.401/ x^2$ or $\Delta I_\nu/x \propto x^2 M \simeq -1.401$.
This mismatch can be partly explained by the fact that in the FH the amplitude of $\mu$ is considered constant, which in reality is not true because of the emission and absorption of photons by BR and DC. Including these effects, one finds \citep{Sunyaev1970mu, Danese1982, Chluba2014}
\begin{equation}
    \mu(\tau,x) \approx \mu_{\infty}(\tau) \cdot {\rm e}^{-\frac{\xc(\tau)}{x}}\,.
\end{equation}
However, given $\xc\simeq \pot{5}{-3}$ for the case considered, this does not fully explain the discrepancy in particular at $x\gtrsim 0.1$ (see Fig.~\ref{fig:fit}). 

We can then ask whether a better match at high frequencies can be achieved by simply fitting the solution within the spectral basis. Performing the fit, we find that 
\[
\begin{array}{c|c}
\mu & 1.37 \times 10^{-5}\,,\\
y & 1.63 \times 10^{-9}\,,\\
\Theta & -5.96 \times 10^{-8}\,,\\
\end{array}
\]
with negligible contributions from $y_{k\geq 1}$ represents the result slightly better (see Fig.~\ref{fig:fit}).
Although the {\tt CosmoTherm} result is photon number conserving (i.e., $\int \Delta N_\nu \id \nu =0$), we see a small improvement by subtracting $G(x)$. This is not surprising, since the lack of photons at $x\lesssim 0.1$ is compensated by a slightly larger distortion amplitude at high frequencies, which requires a subtraction of $G(x)$ to keep the total energy fixed in the fit. 

To improve the FH treatment in a systematic way, one would probably need to consider extending the spectral basis to also include low-frequency emission processes analytically. This is certainly beyond the scope of this paper. However, one 
thing that can be done rather easily is to use the numerical result of {\tt CosmoTherm} for the number density spectral distortion at $\zh=\pot{5}{5}$ and normalize it with the $\mu$ amplitude computed from the distortion visibility function. In this way, we obtain a new numerical spectral shape, $M^*(x)$, as illustrated in the equation below:
\begin{equation}
    \Delta n_{\text{\tt CosmoTherm}} \approx 1.401 \Jdiss(\zh) \Drr \, M^*(x)\,.
\end{equation}
Once computed, $M^*(x)$ can be added to the FH and used instead of the analytical version $M(x)$ to achieve better precision. 

Nevertheless, after these steps we still found a mismatch present at high frequencies, which are fundamental for energetics. This is due to the different scaling of $\nu$ in the high and low regimes. We therefore carried out a further renormalization to fix this difference around the $\mu$-distortion maximum. The final results of our efforts are presented in Fig.~\ref{fig:fhm-5e5-final}, where a significant improvement can be observed in the $\mu$-epoch compared to the other treatments. 
This also emphasizes that for $\zh>\pot{3}{5}$ and $x>0.1$ (or $\nu\gtrsim 20\,\GHz$), {\tt CosmoTherm} is better represented by $M^*(x)$ instead of $M(x)$. In fact, part of the difference comes from frequency-dependent corrections of the interplay between the DC and Compton processes \citep{Chluba2014}. It will be interesting to see if further improvements can be achieved analytically; however, we leave a discussion to the future.

\section{Discussion and Conclusion}
\label{sec:Discussion and Conclusions}
Distortions of the CMB blackbody spectrum can be a useful tracer for pre-recombination physics. However, SDs also carry information about late-time processes; therefore, it is worth studying their evolution at $z<1000$. In Fig.~\ref{fig:lateGF}, we show the {\tt CosmoTherm} results implementing the Green's function method in this redshift range. With a series of $\delta$ functions energy injection rates, we illustrate that at high frequencies, the computed Green's functions are all represented by a $y$-distortion with decreasing amplitude. This is related to the reduced efficiency of the Compton scattering between photons and electrons, as we highlight here. However, the low-frequency part of the relative temperature difference presents an epoch-dependent trend related to free-free processes. Specifically, DC and BR are more efficient in this regime, leading to emission/absorption processes. Moreover, including the Hubble cooling effect to the simulation significantly alters the low-frequency evolution of the distortion, illustrating the sensitivity of the distortion on the electron temperature evolution (see Fig.~\ref{fig:DT-hc}). 

Next, we considered the distortion visibility function at various redshifts. Fig.~\ref{fig::Jbb} compares the numerical result with the analytical function $\Jdc$ computed in previous studies. It was already known that $\Jdc$ overestimates the {\tt CosmoTherm} one for $z_h \gtrsim 10^6$. However, here we highlight the behavior after recombination, where a reduced fraction of the energy reaches the CMB. Moreover, we illustrate how the numerical distortion visibility function is also affected by Hubble cooling, through the time scale for energy transfer and the actual fraction of injected energy that heats the medium. In fact, not all the energy goes into heating; a significant part also serves to excite and ionize the IGM. Our results bracket the range expected for various scenarios.

Finally, in Sect.~\ref{sec:FH_treatment}, we study the differences between the FH treatment and the full {\tt CosmoTherm} result. Our goal was to improve the agreement during the $\mu$-era. As a first step, we investigated whether the amplitude of the Gaussian used to model a $\delta$-energy injection can influence the results, finding this effect to be too small (see Fig.~\ref{fig:rw}). Next, we tried to improve the modeling of the thermalization efficiency, which is determined by the critical frequency parameter. By matching directly to {\tt CosmoTherm}, we improved the agreement at high frequencies, which is fundamental for the energetics of the problem. To resolve the remaining discrepancies, we performed a fit of the {\tt CosmoTherm} output at $\zh = 5 \times 10^5$ using the FH basis; however, this revealed only a slight improvement with the unwanted side effect of adding an unphysical temperature shift to the representation. We therefore decided to produce a new spectral template for the $\mu$-era, $M^*(x)$, using the {\tt CosmoTherm} output. With this new definition, an almost perfect agreement for injections at $\zh\gtrsim \pot{3}{5}$ could be achieved (cf. Fig.~\ref{fig:fhm-5e5-final}).
This suggests that further improvements of the FH treatment can be achieved analytically by including new low-frequency distortion basis contributions. We leave an investigation to future work.

\section*{Acknowledgements}
SE acknowledges support from the Erasmus+ Mobility Programme, and hospitality at the University of Manchester. FP acknowledges partial support from the INFN grant InDark and from the Italian Ministry of University and Research (\textsc{mur}), PRIN 2022 `EXSKALIBUR – Euclid-Cross-SKA: Likelihood Inference Building for Universe's Research', Grant No.\ 20222BBYB9, CUP C53D2300131 0006, and from the European Union -- Next Generation EU. FP and ARF acknowledge support from the FCT project ``BEYLA -- BEYond LAmbda" with ref. number PTDC/FIS-AST/0054/2021.

\section{Data availability}
The data underlying this article are available in this article and can be further made available upon request.

{\small
\vspace{-3mm}
\bibliographystyle{mn2e}
\bibliography{Lit
}

\appendix

\end{document}